\documentstyle[aps,psfig]{revtex}
\begin{document}
%\draft 
\title {{Causality in Relativistic Many Body Theory}
\thanks{Supported by the Deutsche Forschungsgemeinschaft (SFB 201)}}
\author{H. Blum and R. Brockmann}
\address{Institut f\"ur Kernphysik, Johannes Gutenberg-Universit\"at, 
D-55099 Mainz, Germany}
%\date{\today}
\maketitle 

\begin{abstract}
The stability of the nuclear matter system with respect to density
fluctuations is examined exploring in detail the pole structure of
the electro--nuclear response functions. Making extensive use of the
method of dispersion integrals we calculate the {\em full} polarization
propagator not only for real energies in the spacelike and timelike
regime but also in the whole complex energy plane. The latter proved
to be necessary in order to identify unphysical causality violating
poles which are the consequence of a neglection of vacuum polarization.
On the contrary it is shown that Dirac sea effects stabilize the nuclear matter
system shifting the unphysical pole from the upper energy plane back to
the real axis. The exchange of strength between these real timelike
collective excitations and the spacelike energy regime is shown to lead 
to a reduction of the quasielastic peak as it is seen in electron 
scattering experiments. Neglecting vacuum polarization one also obtains 
a reduction of the quasielastic peak but in this case the strength is 
partly shifted to the causality violating pole mentioned above which 
consequently cannot be considered as a physical reliable result. 
Our investigation of the response function in the energy region above 
the threshold of nucleon anti-nucleon production leads to another
remarkable result.  Treating the nucleons as point--like Dirac 
particles we show that for {\em any} isospin independent NN-interaction
RPA-correlations provide a reduction of the production amplitude for
${\mathrm p\bar p}$-pairs by a factor 2.
\end{abstract}
%\newpage

\section{INTRODUCTION}

Dirac sea effects are perhaps the most interesting characteristic of
relativistic many body theory. In nuclear matter e.g. the effective
mass of the nucleons and their coupling to the mesons undergo
considerable corrections due to vacuum fluctuations \cite{Serot},\cite{Chin}.
In addition, relativistic many body theory predicts strong corrections
of the properties of hadrons at high nuclear matter densities. 
Unfortunately, up to now it proved to be very difficult to find unambiguous 
experimental evidence for such effects. One method to probe the Dirac sea
is to investigate the production of antiprotons and other antibaryons in 
heavy ion collisions. Such experiments are currently performed by several
groups under quite different kinematic conditions. These range from
subthreshold production of antiprotons in Ni+Ni collisions at 
1.85 GeV/nucleon at SIS (GSI) \cite{Schroter et al.} up to antibaryon 
production in Au+Au collisions at 14.6 GeV/nucleon at AGS (E878 
Collaboration at BNL) \cite{Beavis et al.}
and sulphur--nucleus collisions at 200 GeV/nucleon at SPS (NA35 collaboration
at Cern) \cite{Alber et al.}. Beside the search for evidence of the 
quark--gluon plasma \cite{quark matter} the attention is drawn to possible
changes of the properties of the nucleons and especially the vector 
mesons \cite{Li} -- \cite{Hatsuda} in the nuclear medium.

The presence of a filled Dirac sea of antiparticles, in addition to the
Fermi sea of protons and neutrons, has consequences not only for the static
properties mentioned above, but also for the dynamics of nuclear matter. 
The latter may be probed e.g. by measuring
the response of the system to the disturbance induced by an external
virtual photon, as it is done in inelastic electron scattering experiments.
A well known and as yet unsolved problem in this field is an apparent
quenching (in comparison with the prediction of simple one--particle models)
of the longitudinal contribution to the inclusive scattering
cross section, particularly for medium heavy and heavier nuclei like $^{40}$Ca,
$^{56}$Fe and $^{238}$U. First observed for momentum transfers of some
100\,MeV/c \cite{Meziani}, \cite{Blatchley}, a SLAC experiment \cite{Chen} 
later showed that the quenching
even persists up to 1\,GeV/c. This may be taken as a hint that relativistic
effects play a role. Further, it seems reasonable that in heavier nuclei 
collective phenomena reduce the strength of longitudinal one--particle 
excitations. So the quenching problem of the longitudinal response function
appears as an appropriate subject for the application of relativistic 
many body theory.

The first conserving approximation which also includes collective
phenomena in a proper way is the random phase approximation (RPA).
RPA--calculations of the response functions for inelastic
electron scattering have been done by several authors \cite{Kurasawa1} -- \cite{Horowitz2}. In the
framework of Walecka's $\sigma\omega$--model \cite{Serot} the first calculations 
were performed for nuclear matter \cite{Kurasawa1} -- \cite{Horowitz1} and then applied to finite nuclei
via the local density approximation (LDA) \cite{Wehrberger1},
\cite{Wehrberger2}. As a result of
many body correlations, in RPA the longitudinal response function is reduced
by about 10--20\% compared to the Mean Field (MFA) and Hartree 
approximation (HA). This 
reproduces at least qualitatively the trend of the experimental data.
Similar results were obtained in a finite nucleus 
calculation \cite{Horowitz2}. 

Despite the final curves of \cite{Kurasawa1} -- \cite{Horowitz2} seem to 
confirm the same qualitative 
picture, they are obtained from two completely different approximation
schemes. While vacuum polarization was neglected in \cite{Kurasawa1} and  
\cite{Wehrberger1}, it is included
in the other calculations. In view of the results for the response
functions one could argue that these seem {\em }not to be very sensitive 
with respect 
to Dirac sea effects. But this is not the case. The true situation is veiled
by the fact that all calculations were performed only for the space--like
kinematic region ($q^{2} = q_{0}^{2} - \vec q^{\,2} < 0$). The reason therefore
is of course that only this regime is accessible by electron
scattering. 

The aim of the present paper now is to draw attention to the fact, 
that effects in the space-- and
time--like energy regimes, which seem to be so well separated in view of the
completely different experimental methods of investigation sketched above,
nevertheless have an influence on each other. Especially it is our goal
to show, that calculations which are performed to explain experimental data
in one of the two kinematic regions have at least to be checked, whether
for the other regime these provide results, which are physically consistent 
at least within the model used. So it will be shown that in the 
$\sigma\omega$--model only calculations including Dirac sea effects 
will preserve causality. 

However, this becomes obvious only if one looks at the time--like part of the 
electro--nuclear response function. In RPA one would expect to find there a 
sharp peak corresponding to the $\sigma$-- and the $\omega$--meson 
respectively, which are the transmitter of the interaction in the Walecka 
model. Through their interaction with the nucleons these acquire the character
of `dressed'  quasi--particles. This takes place in form of repeated 
creation and decay of virtual particle--hole and (if the Dirac sea is also
taken into account) particle--antiparticle pairs which leads to a collective
excitation of the whole nuclear matter system. Because a photon, emitted
by an electron in a scattering process, interact with the nucleonic system 
via the same mechanism, these collective excitations manifest themselves
as peaks in the corresponding response functions. More specific, there are
four degrees of freedom: one scalar from the $\sigma$ and a longitudinal
as well as two transverse from the $\omega$, where the latter are degenerated
because of rotational invariance. Accordingly, two peaks are expected in
the longitudinal response function (because of a mixing between the scalar
and the longitudinal degree of freedom) and one in the transverse branch.
These collective excitations in the time--like energy region have a direct 
back--coupling to the quasielastic bump in the space--like regime: they draw
off strength, which causes the RPA--effect observed in former calculations
\cite{Kurasawa1} -- \cite{Horowitz2}. In addition, these collective excitations
should influence also the nucleonic properties which are studied in the
heavy ion collision experiments mentioned at the beginning of this section.
For example the masses of the mesons as well as the nucleons are shifted
by the interaction \cite{Saito} which has consequences also for the 
subthreshold antiproton production \cite{Li_Ko}. This shows again how deeply
connected phenomena in the two different kinematic regions are.

The mechanism for the reduction of the quasielastic bump, described so far,  
of course makes sense only if the time--like collective excitations which
draw off the strength from the space--like regime are
{\em physical} degrees of freedom. 

As will be shown, this is only the case if 
Dirac sea effects are properly taken into account. A calculation where
vacuum polarization is neglected shows
only one meson peak in the longitudinal and no peak at all in the transverse
branch of the electro--nuclear response function. Instead we find
two peaks in the {\em upper complex energy plane} which means that these
are not corresponding to well defined quasi particles but announce an
instability of the nuclear matter system. Therefore vacuum polarization
is crucial for the stability of nuclear matter.

The purpose of the present paper is to study general properties of 
collective excitations in the nuclear matter system which are rather
a consequence of many body effects then of the basic NN--potential actually
used. Because it is not our intent to compare our curves with experimental
data we use with the Walecka model a simple but nevertheless nontrivial
NN--interaction. This has the advantage that for space-- and time--like (real)
energies the calculations may be done analytically which makes it easier
to perform the numerical extension to complex energies. 

In section II the criteria for the stability of collective excitations will be
discussed in general. In order to explore the upper complex energy plane
with respect to destabilizing unphysical modes we have to perform an analytic
continuation of the electro--nuclear response functions. This will be done
in section III. Using dispersion relations we introduce a renormalization
concept \cite{Berestetskii} which is new in this context. Instead of using 
dimensional regularization -- as it was done in \cite{Kurasawa2} and 
\cite{Wehrberger2} -- we perform subtractions in the dispersion integrals 
for the polarization tensor in order to get rid of the infinities due to
vacuum polarization. We show explicitly that both renormalization technics
lead to the same results for the response functions which is an interesting
result in itself.

In section IV the response functions are at first discussed for real energies.
In section V we look at the single particle properties in the space--like
(quasielastic bump) and the time--like regime (particle--antiparticle 
excitations). In both cases as a consequence of RPA--correlations we find 
a reduction of the response functions compared to the independent particle 
model. This effect is most
striking for the particle--antiparticle production amplitude which for high
energies is reduced exactly by a factor 2. We show that this halving
of the amplitude is a general result which holds true for {\em any}
isospin independent interaction. This is one of the main results of the 
present paper.

In section VI we discuss the role of collective excitations extending our
considerations from real energies to the upper complex energy plane. 
Neglecting vacuum polarization we find the result mentioned already above,
that the nuclear matter system becomes unstable because of excitations
with complex energy $z$ where ${\mathrm Im}(z) > 0$. Finally our conclusions 
are given in section VII.

\section{STABILITY OF COLLECTIVE EXCITATIONS}

A possible mass change of light vector mesons inside nuclear
matter as a consequence of medium modifications of the quark condensate
$<\bar qq>$ \cite{Li} has recently drawn attention to calculations
of timelike collective excitations in various models. An comprehensive 
overview of this work can be found in \cite{Hatsuda} and the references
given therein. Hatsuda et. al. estimate the mass reductions of 
$\rho$, $\omega$ and $\Phi$ using QCD sum rules and compare these
results with a calculation in the framework of the $\sigma\omega$--model
(which is suitable extended in order to include strangeness and the 
$\Phi$--meson). In the latter approach which is similar to ours
in this paper they obtain the effectives masses from zeroes
of the inverse meson propagator in the long wavelength limes 
$|\vec q| \rightarrow 0$ inside nuclear matter.  

Apart from this actual interest in density dependent changes of the
quark condensate collective modes in nuclear matter were studied 
already previously by several groups. The first calculation within the
$\sigma\omega$--model was performed by Chin \cite{Chin}.
Neglecting vacuum polarization he did an additional severe approximation
\footnote{As a result of this approximation, beside vacuum polarization
also the density dependent Pauli blocking of ${\mathrm N\bar N}$ excitations 
due to the filled Fermi sea is excluded from the calculation.}
evaluating the momentum integrals in the expression for the dielectric 
function. So Chin's results are partly in contradiction to the work
of Lim and Horowitz \cite{Lim} who calculated the density dependent momentum
integrals exactly but neglected vacuum polarization too. These authors
also addressed the problem of the stability of collective modes, however
from a different point of view then ours in this paper. Lim and Horowitz
investigated the density dependence of {\em real} collective zero energy 
modes ($q_{0}=0$) in the space--like region. If for a given density
(characterized by the Fermi momentum $k_{F}$) and momentum $|\vec q|$
such a mode arises, this indicates that in the uniform nuclear matter
system an instability occurs which drives a phase transition to a
spatial inhomogeneous state. A similar type of instability was studied
by Furnstahl and Horowitz \cite{Furnstahl} in connection with the so called
Landau ghost which appears in meson propagators when vacuum polarization
is included. Neither of these singularities will affect our calculations.
The first type is excluded because we are only interested in nuclear
matter at saturation density ($k_{F} = 1.4\,fm^{-1}$) where the system
is stable against zero energy modes (see fig.\,12 of \cite{Lim}).
The Landau ghost appears only for momentum transfers of 
$|\vec q|\geq 2.55\,m$ ($m=$ nucleon mass) \cite{Furnstahl} which is 
much higher then
the momenta we are considering 
\footnote{In a paper by Tanaka et al. \cite{Tanaka} it was shown that the
Landau ghost is not a serious problem of relativistic many body theory
because it can be eliminated in an appropriate way}.

In the following we examine the nuclear matter system with respect to
a different kind of instabilities. In the previous calculations of 
collective nuclear matter excitations mentioned above, only the denominator
of the meson propagator was analyzed on the real energy axes. Zeroes of
this denominator are good candidates for corresponding poles in the 
propagator i.e. an excitation of the system. But, while even a zero of
real {\em and} imaginary parts of the denominator is not always a sufficient 
condition for a pole (if the nominator also becomes small), often only 
the vanishing of the real part for an given energy $\omega_{0}$ is already 
taken for an excitation. If the imaginary part of the denominator keeps 
sufficient small (and if the nominator is finite) this is correct and 
the propagator in the neighbourhood of $\omega_{0}$ has the shape of a
Lorentz curve corresponding to an excitation with a decay width depending
on the magnitude of the imaginary part.  

In a paper by Jean et. al. \cite{Jean} the denominator of the $\omega$-meson
propagators has been analyzed similar to \cite{Shiomi} and \cite{Hatsuda} in order 
to find a mass shift as result of medium modifications in the 
$\sigma\omega$--model. As it is one of the aims of the present paper, 
Jean et. al. also studied the influence of the Dirac sea
on the effective $\omega$ mass comparing the full calculation with
one where vacuum polarization is neglected and only Pauli blocking is
taken into account. While in the presence of vacuum polarization
all degrees of freedom are correctly reflected by a corresponding pole
in the propagator they found in the latter case zeroes of the real part
of the denominator simultaneously with a nonvanishing {\em negative}
imaginary contribution. They concluded that "this nonzero imaginary
part contributes to an unphysical (i.e. negative) decay width for the
$\omega$ meson".

As mentioned above, at this point it becomes obvious that it is not 
sufficient to look only at the denominator of the propagator. In 
section VI we will show that the zero of the real part of the latter,
found by Jean et. al., does not correspond to a collective excitation
with "negative decay width" (whatever this means) at least not on
the real energy axes. However, in order to find out what is really
hidden behind this unphysical mode, it is necessary to explore
the full propagator (not only it's denominator) in the whole
complex energy plane. As noticed already in the introduction,
the neglect of vacuum polarization entails an instability of
the nuclear matter system indicated by poles of the 
electro--nuclear response functions in the upper complex energy 
plane. 

In the case of inclusive (e,e')--scattering there are
two independent response functions, a longitudinal and a transverse, which
can be expressed by the corresponding matrix elements of the polarization 
tensor $\Pi^{\mu\nu}(q_{0},\vec q)$. If we choose a co--ordinate system
were $\vec q$ is parallel to the $x$--axes,  $\Pi^{\mu\nu}(q_{0},\vec q)$ 
can be written in the form:
\begin{eqnarray}\label{0}
\Pi^{\mu \nu}(q_{0},\vec q) & = &
\left(
\begin{array}{cccc}
\Pi_{\mathrm L}(q_{0},\vec q) & \frac{q_{0}}{|\vec q|}
\Pi_{\mathrm L}(q_{0},\vec q)& 0 & 0 \\
\frac{q_{0}}{|\vec q|}\Pi_{\mathrm L}(q_{0},\vec q) &
\left(\frac{q_{0}}{|\vec q|}\right)^{2}\Pi_{\mathrm L}(q_{0},\vec q) & 0 & 0 \\
0 & 0 &\Pi_{\mathrm T}(q_{0},\vec q) & 0 \\
0 & 0 & 0 & \Pi_{\mathrm T}(q_{0},\vec q)
\end{array}
\right)\;.
\end{eqnarray}
The Fourier transform of $\Pi^{\mu\nu}(q_{0},\vec q)$:
\begin{equation}\label{1} 
\Pi^{\mu\nu}(x,y) = \int\limits_{-\infty}^{+\infty}\,\frac{dq_{0}}{2\pi}
\int\,\frac{d\vec q}{(2\pi)^{3}}\,\Pi^{\mu \nu}(q_{0},\vec q)
e^{-iq_{0}(t_{x}-t_{y})}e^{i\vec q(\vec x - \vec y)}
\end{equation}
is defined as the change $\delta j^{\mu}(x)$ of the nuclear current in $x$
due to a change $\delta A_{\nu}(y)$ of the vector potential in $y$, induced
by the scattered electron:
\begin{equation}\label{2}
\delta j^{\mu}(x) = \Pi^{\mu\nu}(x,y)\,\delta A_{\nu}(y)\;.
\end{equation}
According to linear response theory \cite{Fetter} $\Pi^{\mu\nu}(x,y)$ can be
expressed by the ground state expectation value of the commutator 
$\left[j^{\mu}(x),j^{\nu}(y)\right]$.
Causality requires that $\Pi^{\mu\nu}(x,y)$ can be non zero only if
$x$ is in the forward light cone with respect to $y$, i.e. $t_{x} > t_{y}$.
Including a factor (-i) which is conventional, $\Pi^{\mu\nu}(x,y)$ can
be written explicitly:
\begin{equation}\label{2a}
\Pi^{\mu\nu}(x,y)= -i\,<0\left|\,\left[j^{\mu}(x),j^{\nu}(y)\right]\,
\right|0>\,\Theta(t_{x}-t_{y})\;.
\end{equation}
Here $|0>$ denotes the ground state of the nuclear matter system 
consisting of a filled Fermi sea of nucleons above the Dirac vacuum of
antiparticles.

Now, for $t_{y} > t_{x}$ the $q_{0}$--integral in (\ref{1}) has to be closed in
the upper complex energy plane. Then a zero contribution can only be obtained
if $\Pi^{\mu \nu}(q_{0},\vec q)$ is analytic there. On the contrary, if 
$\Pi^{\mu \nu}(q_{0},\vec q)$ has poles in the upper half plane, this means
that there is a response of the system preceding an external disturbance
which is equivalent to an instability. Because in \cite{Chin},
\cite{Lim}, \cite{Furnstahl} only poles
on the real $q_{0}$--axis were examined, non causal singularities could not
be found by these authors. 

\section{ANALYTIC CONTINUATION TO COMPLEX ENERGIES}

For nuclear matter (point--like Dirac particles)
with an interaction mediated by a scalar and a vector
meson ($\sigma\omega$--model \cite{Serot}) the longitudinal and transverse 
electro--nuclear response functions within RPA are given by the following
expressions \cite{Kurasawa1}
\footnote{Note that in contrast to (\ref{2a}) in \cite{Kurasawa1} 
$\Pi^{\mu\nu}(x,y)$ 
is defined as the time ordered product \\
$-i\,<0\left|\,T\,j^{\mu}(x),j^{\nu}(y)\,\right|0>$. For $q_{0} > 0$
the Fourier transform of this expression is identical to our 
$\Pi^{\mu \nu}(q_{0},\vec q)$ and differs only in the sign of the imaginary
part for  $q_{0} < 0$.}:
\begin{eqnarray}\label{3}
& & \Pi_{{\mathrm RPA}}^{\mathrm L}(q_{0},\vec q^{\,2}))\;
=\; \frac{e^{2}}{2}\Pi_{vv}^{\mathrm L}(q_{0},\vec q^{\,2}) \nonumber\\
& &\,+\, e^{2}\, \frac{(1-\frac{q^{2}}{\vec q^{\,2}}G_{v}
\Pi_{vv}^{\mathrm L})G_{s}\left(\frac{1}{2}\Pi_{sv}\right)^{2}
\,+\,\frac{1}{2}\frac{q^{2}}{\vec q^{\,2}}G_{v}G_{s}\Pi_{vv}^{\mathrm L}
\Pi_{sv}^{2}
\,+\,\frac{q^{2}}{\vec q^{\,2}}(1-G_{s}\Pi_{ss})
G_{v}\left(\frac{1}{2}\Pi_{vv}^{\mathrm L}\right)^{2}}
{(1-\frac{q^{2}}{\vec q^{\,2}}G_{v}\Pi_{vv}^{\mathrm L})
(1-G_{s}\Pi_{ss})
\,-\,\frac{q^{2}}{\vec q^{\,2}}G_{v}G_{s}\Pi_{sv}^{2}} \;, 
\nonumber\\
\end{eqnarray}
\begin{eqnarray}\label{4}
\left[\Pi_{{\mathrm RPA}}\right]_{\mathrm T}(q_{0},\vec q^{\,2}) & = & 
\frac{e^{2}}{2}\Pi_{vv}^{\mathrm T}(q_{0},\vec q^{\,2}) + 
e^{2} \frac{G_{v}(q^{2})\left(\frac{1}{2}
\Pi_{vv}^{\mathrm T}(q_{0},\vec q^{\,2})\right)^{2}}
{1\,-\,G_{v}(q^{2})\Pi_{vv}^{\mathrm T}(q_{0},\vec q^{\,2})} \;,
\end{eqnarray}
with:
\begin{eqnarray}\label{5}
G_{s,v}(q^{2}) & = & \frac{g_{s,v}^{2}}{q^{2}-m_{s,v}^{2}+i\epsilon} \;.
\end{eqnarray}
$e, g_{s}$ and $g_{v}$ are the electromagnetic, scalar and vector meson
coupling constant respectively. $\Pi_{ab}(q_{0},\vec q)$ ($a,b= s,v$) 
is the Amplitude for a scalar or vector boson ``a'' to go into a 
nucleon--hole or nucleon--antinucleon pair which recombines to a boson ``b''
\footnote{For vector bosons $\Pi_{vv}\equiv \Pi_{vv}^{\mu\nu}$ 
is a $4\times4$--tensor, the longitudinal and transverse
matrix elements of which are denoted by $\Pi_{vv}^{\mathrm L}$ and
$\Pi_{vv}^{\mathrm T}$. Because we have drawn out the coupling constants
and isospin factors from the vertices, the analytical form of $\Pi_{vv}$ 
is the same for  $\omega$--mesons as for photons. A difference between 
$\Pi_{\omega\omega}$, $\Pi_{\omega\gamma}$ and $\Pi_{\gamma\gamma}$ will
occur only in the numerical expressions of the vacuum contributions as
a consequence of the renormalization procedure (see (\ref{19})).}. 
In former nuclear matter calculations \cite{Kurasawa1} -- \cite{Horowitz1}
$\Pi_{ab}(q_{0},\vec q)$
was 
evaluated explicitly for real $q_{0}$ employing Feynman rules. In order to 
get a finite vacuum contribution dimensional regularization was performed.
In the following we use a method developed by Berestetskii in the context
of QED \cite{Berestetskii} which is more appropriate for
an analytic continuation of (\ref{3}) and (\ref{4}) to complex energies
$z:=q_{0}+iq'_{0}$. We start from the imaginary parts of $\Pi_{aa}$ for
energy transfers $q_{0} \ge 0$.
Unitarity requires that these are given by the probability of boson ``a''
going into a nucleon--hole or a nucleon--antinucleon pair:
\begin{equation}\label{7}
{\mathrm Im}\,\Pi_{ss}(q_{0},\vec q^{\,2}) = (4m^{*\,2}-q^{2})I_{0}(q_0,\vec q^{\,2}) 
+ {\mathrm Im}\,\Pi_{ss}^{vac}(q^2,m^{*\,2})\;\;;
\end{equation}
\begin{equation}\label{8}
{\mathrm Im}\,\Pi_{vv}^{\mathrm L}(q_{0},\vec q^{\,2}) = 
-\vec q^{\,2}I_{0}(q_0,\vec q^{\,2})
+I_{2}(q_0,\vec q^{\,2}) + \vec q^{\,2}{\mathrm Im}\,\Pi_{vv}^{vac}
(q^2,m^{*\,2})\;\;;
\end{equation}
\begin{equation}\label{9}
{\mathrm Im}\,\Pi_{vv}^{\mathrm T}(q_{0},\vec q^{\,2}) = 
-(4m^{*\,2}+q^{2})I_{0}(q_0,\vec q^{\,2})
-\frac{q^{2}}{\vec q^{\,2}}I_{2}(q_0,\vec q^{\,2}) - 
q_{0}^{2}{\mathrm Im}\,\Pi_{vv}^{vac}(q^2,m^{*\,2})\;\;.
\end{equation}
The density dependent contributions to these amplitudes are given by the
integrals $I_{n}$ over Fermi sea momenta up to the Fermi edge 
$k_{\mathrm F}$:
\begin{eqnarray}\label{10}
I_{n}(q_{0},\vec q^{\,2})& = -&\frac{1}{4\pi^{2}}\int \,d^{3}k
\frac{n_{\vec k}}{\epsilon_{\vec k}\epsilon_{\vec k + \vec q}}\left\{
(1-n_{\vec k + \vec q})(2\epsilon_{\vec k}+q_{0})^{n}
\delta(q_{0}-[\epsilon_{\vec k + \vec q}-\epsilon_{\vec k}])\right.
\nonumber \\
& &\qquad\qquad + \left.(2\epsilon_{\vec k}-q_{0})^{n}
\delta(q_{0}-[\epsilon_{\vec k + \vec q}+\epsilon_{\vec k}])\right\}
\;\;.
\end{eqnarray}
Here $n_{\vec k} = \Theta(k_{\mathrm F}-|\vec k|)$ denotes the Fermi 
distribution of the mean field ground state and 
$\epsilon_{\vec k} = \sqrt{\vec k^{2}+m^{*\,2}}$ is the energy of a nucleon
with wave vector $\vec k$ and effective mass $m^{*}$. The 
$\delta$--functions in the integrand of (\ref{10}) make sure that the 
two terms give a contribution only in the space--like or the time--like region
respectively. The momentum integral can be easily performed analytically 
\cite{Kurasawa1}.

The Dirac sea contributions of boson ``a'' going into a 
nucleon--antinucleon pair are given by:
\begin{equation}\label{11}
{\mathrm Im}\,\Pi_{ss}^{vac}(q^2,m^{*\,2}) = -\frac{q^{2}}{4\pi}
\left(1-\frac{4m^{*\,2}}{q^{2}}\right)^{\frac{3}{2}}\Theta(q^{2}-4m^{*\,2})
\end{equation}
and
\begin{equation}\label{12}
{\mathrm Im}\,\Pi_{vv}^{vac}(q^2,m^{*\,2}) = -\frac{1}{12\pi}
\left(1-\frac{4m^{*\,2}}{q^{2}}\right)^{\frac{1}{2}}
\left(1+\frac{2m^{*\,2}}{q^{2}}\right)
\Theta(q^{2}-4m^{*\,2})\;\;.
\end{equation}

In addition to the ``proper'' transitions 
$a \rightarrow ($ph or ${\mathrm N\bar N}) \rightarrow a$
($a= s,v$) there is also a mixing of scalar and vector degrees of freedom.
However, this is a purely density dependent effect. Because the vector
current of the vacuum is zero, its commutator with the scalar current 
(which according to (\ref{2a}) enters the expression 
for the components $\Pi_{sv}$ of the polarization
tensor) vanishes. The vector current of the nuclear matter system on the 
other side has a non--vanishing time--like component which is given by
the energy density of the filled Fermi sea. The corresponding
matrix element of the polarization tensor for $q_{0} \ge 0$ has the form:
\begin{equation}\label{13}
{\mathrm Im}\,\Pi_{sv}(q_{0},\vec q^{\,2}) = 
2m^{*\,2}I_{1}(q_{0},\vec q^{\,2})\;.
\end{equation}
For $q_{0} < 0$ ${\mathrm Im}\,\Pi_{ab}(q_{0},\vec q^{\,2})$ is given by the 
negative of the expressions (\ref{7}), (\ref{8}), (\ref{9}) and (\ref{13})
respectively:
\begin{equation}\label{14}
{\mathrm Im}\,\Pi_{ab}(q_{0},\vec q^{\,2}) = -{\mathrm Im}\,\Pi_{ab}(|q_{0}|,\vec q^{\,2})\qquad
(q_{0} < 0)\;,
\end{equation}
i.e. the spectral functions ${\mathrm Im}\,\Pi_{ab}(q_{0},\vec q^{\,2})$ are 
antisymmetric with respect to $q_{0}$. 

Making use of the spectral representation:
\begin{eqnarray}\label{15}
\Pi_{ab}(z,\vec q^{\,2}) & = & 
\int \limits_{-\infty}^{\infty}\frac{dq_{0}}{\pi}
\frac{{\mathrm Im}\Pi_{ab}(q_{0},\vec q^{\,2})}{q_{0}-z} 
\end{eqnarray}
from the analytic expressions for the imaginary parts (\ref{7})--(\ref{9}) 
and (\ref{13}) the full $\Pi_{ab}$ can be obtained in the whole 
complex energy plane. For $q_{0} \ge 0$, ${\mathrm Im}\,\Pi_{ab}$ can be treated
as a function of $q_{0}^{2}$. Using the antisymmetry property 
(\ref{14}) we can write (\ref{15}) also in the form:
\begin{eqnarray}\label{16}
\Pi_{ab}(z^2,\vec q^{\,2}) & = & 
\int \limits_{0}^{\infty}\frac{dq_{0}^{2}}{\pi}
\frac{{\mathrm Im}\Pi_{ab}(q_{0}^{2},\vec q^{\,2})}{q_{0}^{2}-z^{2}}\;. 
\end{eqnarray}
However (\ref{16}) is directly applicable only
to the density dependent portion of $\Pi_{ab}$ because this has the
appropriate asymptotic behaviour for $q_{0} \rightarrow \infty$.
In the high energy regime the imaginary parts of $\Pi_{ss}$ and $\Pi_{vv}$
are given by the vacuum contributions ${\mathrm Im}\,\Pi_{ss}^{vac}(q^2)$ and
${\mathrm Im}\,\Pi_{vv}^{vac}(q^2)$, the asymptotic behaviour of which for
$q_{0}\rightarrow \infty, (\vec q^{\,2} = const.)$ follows from
(\ref{11}) and (\ref{12}):
\begin{eqnarray}\label{17}
{\mathrm Im}\,\Pi_{ss}^{vac}(q_{0}^{2},\vec q^{\,2})& \propto & q_{0}^{2}
\nonumber \\
{\mathrm Im}\,\Pi_{vv}^{vac}(q_{0}^{2},\vec q^{\,2})&\propto& const. 
\end{eqnarray}
Therefore in order to get from (\ref{16}) a finite result for the vacuum 
contributions, we have to perform in the dispersion integrals one subtraction 
with respect to $q_{0}^{2}$ for $\Pi_{vv}$ and two subtractions for $\Pi_{ss}$:
\begin{eqnarray}\label{18}
\Pi_{vv}^{vac}(z^{2},\vec q^{\,2}) & = & 
\Pi_{vv}^{vac}(z_{r;v}^{2},\vec q^{\,2}) +
(z^{2}-z_{r;v}^{2})\int \limits_{0}^{\infty}\frac{dq_{0}^{2}}{\pi}
\frac{{\mathrm Im}\Pi_{vv}^{vac}(q_{0}^{2},\vec q^{\,2})}{(q_{0}^{2}-z_{r;v}^{2})
(q_{0}^{2}-z^{2})} \; , \nonumber \\
\Pi_{ss}^{vac}(z^{2},\vec q^{\,2}) & = & 
\Pi_{ss}^{vac}(z_{r;s}^{2},\vec q^{\,2}) +
\left.(z^{2}-z_{r;s}^{2})\frac{d\,\Pi_{ss}^{vac}(z^{2},\vec q^{\,2})}
{dz^{2}}\right|_{z^{2}=z_{r;s}^{2}} \nonumber\\
& & + (z^{2}-z_{r;s}^{2})^{2}\int \limits_{0}^{\infty}\frac{dq_{0}^{2}}{\pi}
\frac{{\mathrm Im}\Pi_{ss}^{vac}(q_{0}^{2},\vec q^{\,2})}
{(q_{0}^{2}-z_{r;s}^{2})^{2}(q_{0}^{2}-z^{2})} \;.
\end{eqnarray}
Using (\ref{17}) it can be easily checked that the integrals in (\ref{18})
are indeed finite. This procedure is
equivalent to the corresponding subtraction of infinities in the 
dimensional regularization method \cite{Kurasawa2}, \cite{Horowitz1} or any other renormalization
scheme. Accordingly, the finite values of $\Pi_{vv}^{vac}$, $\Pi_{ss}^{vac}$ and 
$d\,\Pi_{ss}^{vac}/d\,z^{2}$ have to be fixed at appropriate
renormalization points $z_{r}^{2}$.  We have chosen to follow the 
``on mass shell'' renormalization of \cite{Kurasawa2} 
\footnote{There is a certain arbitrariness in the renormalization procedure 
for {\em effective} field theories because it can be argued that the 
masses of the scalar and vector bosons, mediating the interaction in 
these theories, do not have to be identical with the masses of physical
mesons. Thus some authors \cite{Horowitz1}, \cite{Furnstahl} prefer to 
take e.g. $q^{2}=0$ as 
renormalization point . Fortunately the numerical results for the response 
functions are only
%not changed substantially
slightly changed by this different choice.}:
\begin{eqnarray}\label{19}
\Pi_{\gamma\gamma}^{vac}(z_{r;\gamma}^{2}=\vec q^{\,2},m^{*}=m) & = & 0
\nonumber\\
\Pi_{\omega\gamma}^{vac}(z_{r;\omega}^{2}=\vec q^{\,2}+m_{\omega}^{2},
m^{*}=m) & = & 0
\nonumber\\
\Pi_{\omega\omega}^{vac}(z_{r;\omega}^{2}=\vec q^{\,2}+m_{\omega}^{2},
m^{*}=m) & = & \left. -m_{\omega}^{2}
\frac{d\,\Pi_{\omega\omega}^{vac}(z^{2},m^{*}=m)}{d\,z^{2}}
\right|_{z_{r;\omega}^{2}=\vec q^{\,2}+m_{\omega}^{2}} 
\;\equiv\;\delta m_{\omega}^{2}
\nonumber\\
\Pi_{\sigma\sigma}^{vac}(z_{r;\sigma}^{2}=\vec q^{\,2}+m_{\sigma}^{2},
m^{*}=m) & = & \left. 
\frac{d\,\Pi_{\sigma\sigma}^{vac}(z^{2},m^{*}=m)}{d\,z^{2}}
\right|_{z_{r;\sigma}^{2}=\vec q^{\,2}+m_{\sigma}^{2}}
\;=\; 0 \;.
\end{eqnarray}
Here $\delta m_{\omega}^{2}$ denotes a mass shift term which is introduced
in order to keep the pole of the renormalized $\omega$--propagator in
the vacuum at the known mass of $m_{\omega}=783$\,MeV. 
With the boundary conditions (\ref{19}) the (finite) vacuum contributions 
to the response functions (\ref{18}) are completely determined in the whole 
complex energy plane.

\section{RESPONSE FUNCTIONS}

Having obtained explicit expressions for $\Pi_{ab}(z^{2},\vec q^{\,2})$,
the analytic continuation of the RPA--expressions (\ref{3}) and (\ref{4})
can be easily obtained by the substitution $q_{0}\rightarrow 
z=q_{0}+iq_{0}'$. On the contrary, the original expressions (\ref{3}) and (\ref{4})
can be recovered from $\Pi_{\mathrm RPA}(z,\vec q^{\,2})$ 
\footnote{The same holds true of course for $\Pi_{ab}$ according to 
(\ref{18})}
by the prescription:
\begin{equation}\label{20}
\Pi_{\mathrm RPA}(q_{0},\vec q^{\,2}) = 
\Pi_{\mathrm RPA}(z=q_{0}+i\epsilon,\vec q^{\,2}).
\end{equation}
Here it should be noted that the extrapolation to {\em complex} energies 
(\ref{16}) is even 
necessary to obtain the correct result for the response function on the 
{\em real} energy axis. The reason is, that the $\delta$--like peaks in 
${\mathrm Im}\,\Pi_{\mathrm RPA}$ which are caused by the collective 
excitations may be 
calculated only taking the proper limes according to (\ref{18}) (see 
section VI).

The numerical values of the parameters of the theory are given in
\cite{Serot} :
\begin{equation}\label{21}
\begin{array}{ccc}
m_{s}=550\,{\mathrm MeV}      & \qquad\qquad  &  m_{v}=783\,
{\mathrm MeV} \nonumber\\
[.4cm]
g_{s}^{2}= \left\{
\begin{array}{ll}
92.87 & ({\mathrm MFA})\\
69.87 & ({\mathrm HA})
\end{array} \right.& \qquad\qquad &
g_{v}^{2}= \left\{
\begin{array}{ll}
135.7& ({\mathrm MFA})\\
79.8 & ({\mathrm HA})
\end{array} \right.
\end{array}
\end{equation}
The coupling constants are adjusted to the saturation property of nuclear
matter at an energy of $-16\,{\mathrm MEV}$ per nucleon for a density 
corresponding
to a Fermi momentum of $k_{\mathrm F} = 280\,{\mathrm MEV}/c$. While in mean 
field
approximation (MFA) vacuum effects are neglected, the Hartree 
approximation
includes these properly. The corresponding values for the effective mass
of the nucleon are given by $m^* = 0.56\,m$ (MFA) and $m^* = 0.78\,m$
(HA) respectively. 

Performing the limiting procedure according to (\ref{20}) the 
electro--nuclear response functions can be obtained from 
$\Pi(z,\vec q^{\,2})$:
\begin{equation}\label{22}
{\mathrm R}_{\mathrm L,T}(q_{0},\vec q^{\,2}) 
= -\frac{V}{\pi}\,{\mathrm Im} \Pi_{\mathrm L,T}(q_{0}+i\epsilon,\vec q^{\,2}).
\end{equation}

\section{SINGLE PARTICLE PROPERTIES FOR SPACE-- AND TIMELIKE MOMENTUM 
TRANSFERS}
Part a) of figs.\,1-4 show ${\mathrm R}_{\mathrm L}$ and ${\mathrm R}_{\mathrm T}$ for a 
momentum transfer of $|\vec q|=1.5\,k_{\mathrm F}$ which was a typical value 
in former electron scattering experiments \cite{Meziani}, \cite{Blatchley}.
Part b) of the figures show the corresponding real portions 
 ${\mathrm Re}\,\Pi_{\mathrm L}$ and ${\mathrm Re}\,\Pi_{\mathrm T}$ of the longitudinal
and transverse matrix elements of the polarization tensor
(see equ. (\ref{0})). The energy ranges
from the space--like region 
($q_{0}<|\vec q|$) up to the deep time--like regime ($q_{0}>|\vec q|$).
While vacuum polarization is included in figs.\,3 and 4, in figs.\,1 and 2 
purely density dependent contributions are shown. As mentioned already in 
section I, our goal is to point out qualitative
differences between these two approximations rather then to compare the
results with experimental data (which for the quasielastic regime -- where
such data are only available up to now -- has been done by several authors
\cite{Wehrberger1}, \cite{Wehrberger2}, \cite{Horowitz2}, \cite{Furnstahl}).
Therefore we did not include form factors but performed the calculations
for point--like Fermions in nuclear matter. The dashed curves show the 
results including only one polarization bubble (corresponding to the first
term in equs. (\ref{3}) and (\ref{4}) respectively), while in the full curves 
RPA--correlations are taken into account according to the the second 
expressions in (\ref{3}) and (\ref{4}). The structure on the left sides 
($\omega/k_{\mathrm F} \stackrel{<}{\sim} 1$) of the four figures represents 
the quasielastic bump which is measured in (e,e') scattering experiments. 
According to (\ref{21}) the effective mass $m^{*}$
is higher when vacuum contributions are included. This causes the
quasielastic peak to be a bit narrower in this case \cite{Kurasawa2}, 
\cite{Horowitz1}. The cusp at $\omega/k_{\mathrm F} {\sim} 0.18$
in the 
RPA-curve of fig.\,1a is due to an {\em almost} zero point of the denominator
in equ. (\ref{3}) which indicates that long-range correlations are important. 

From figs.\,1a and 2a one easily reads off that the purely density dependent 
contributions 
to the response functions are zero for $\omega \rightarrow \infty$. Therefore
${\mathrm Re}\,\Pi_{\mathrm L,T}$ in figs.\,1b and 2b can be obtained from the curves in 
Figs.\,1a and 2a directly (i.e. without any subtractions) by means of 
equs. (\ref{16}) and (\ref{22}). 
The quasielastic bump in the response functions
has its counterpart in a `oscillator--structure' in the real parts of 
$\Pi_{\mathrm L,T}$. The cusp at $\omega/k_{\mathrm F} {\sim} 0.18$
in the RPA-curve of fig.\,1a corresponds to a small `shoulder' at the same 
energy in fig.\,1b. 

With vacuum polarization as well as without, many body effects which
are taken into consideration in the RPA provide a considerable reduction
of the longitudinal and a smaller decrease of the transverse 
response functions. The reason for this suppression in the space--like region 
has its origin in the time--like regime and consequently keeps hidden for 
electron scattering experiments: strength is transfered to collective modes 
which in the RPA-curves appear as sharp peaks in the so called `unphysical 
region' $|\vec q| < \omega < \sqrt{4m^{*2}+\vec q^{\,2}}$. 
These peaks will be discussed in more detail below. Despite the RPA-curves
are quite different in shape, the omission of vacuum polarization does not 
change drastically the qualitative picture in the quasi-free regime. Comparing
figs.\,1 and 2 with figs.\,3 and 4 one sees that the last statement holds 
true for
the response functions itself as well as for the the real parts of 
$\Pi_{\mathrm L,T}$.

This changes completely in the time--like energy range. For
$q_{0} \ge \sqrt{4m^{*\,2}+\vec q^{\,2}}$ the production of 
${\mathrm N\bar N}$--pairs from the Dirac sea becomes possible. According to 
(\ref{8}) and (\ref{12}) for the longitudinal response (fig.\,3a)
this vacuum polarization has a ``square root-like''  $q_{0}$-dependence
(but the slope keeps finite) at
threshold and approaches a constant value for increasing energies.
\footnote{Note that the curves in figs.\,3a and 4a show the {\em sum} of
vacuum polarization and the corresponding density dependent contribution.
But it follows from (\ref{7})\,--\,(\ref{10}) (and can be explicitly read off from 
figs.\,1a and 2a)
that in the time--like region the latter are nonzero only for the small energy
range ${\mathrm min}(2\epsilon_{|\vec q|/2},
\epsilon_{\vec k_{\mathrm F} - |\vec q|}+\epsilon_{k_{\mathrm F}}) \leq
q_{0} \leq \epsilon_{\vec k_{\mathrm F} + 
|\vec q|}+\epsilon_{k_{\mathrm F}}$.
So for higher energies in the presence of nuclear matter the shape of the 
curves is the same as for the vacuum.}

From equs. (\ref{8}) and (\ref{9}) one finds for $q_{0} >> |\vec q|$ 
as a relation between the longitudinal and transverse matrix elements of 
${\mathrm Im}\,\Pi_{vv}^{\mu\nu}$:
\begin{equation}\label{23}
\frac{{\mathrm Im}\,\Pi_{vv}^{\mathrm T}(q_{0}^{2},\vec q^{\,2})}
{{\mathrm Im}\,\Pi_{vv}^{\mathrm L}(q_{0}^{2},\vec q^{\,2})} 
\sim \frac{q_{0}^{2}}{\vec q^{\,2}} \sim \frac{q^{2}}{\vec q^{\,2}}\;.
\end{equation}
As will be
shown below, for high energies not only the one bubble contribution but also
the RPA--corrections to $\Pi_{{\mathrm RPA}}^{\mathrm L,T}$
in (\ref{3}) and (\ref{4}) are proportional to $\Pi_{vv}^{\mathrm L,T}$ and
consequently also obey the relation (\ref{23})
\footnote{From the solid and dashed curves in figs.\,3 and 4 one sees already 
that in the high energy limes the two approximations can differ only by a 
multiplicative factor which -- as will be proven analytically below -- 
has the value 1/2. Therefore in the following considerations on the 
asymptotic behaviour of the polarization propagator we do not have to 
distinguish between $\Pi_{{\mathrm RPA}}^{\mathrm L,T}$ and 
$\Pi_{vv}^{\mathrm L,T}$.}.
Hence vacuum contribution 
to ${\mathrm R}_{\mathrm T}$ in fig.\,4a in comparison with ${\mathrm R}_{\mathrm L}$ in fig.\,3a
is weighted by an additional factor $q^{2}$ which causes a rapid increase 
for energies above threshold. So in order to show the low and high 
energy branches in the same figure we had to scale the transverse 
${\mathrm N\bar N}$--contribution by a factor 0.05\,.

In the presence of nuclear matter the production of particle--hole pairs
from the Dirac sea is inhibited if the final state has momentum lower then
$k_{\mathrm F}$. So the Fermi sea causes a reduction of the 
vacuum amplitude for the process $e^{+}e^{-} \rightarrow {\mathrm N\bar N}$. 
Figs.\,1a and 2a show
this negative density dependent contribution to the longitudinal and
transverse response function in the time--like energy regime. The
big slope of ${\mathrm R}_{\mathrm L,T}$ at threshold is reflected as a cusp structure
in the real parts of $\Pi_{\mathrm L,T}$ (see figs.\,1b and 2b).  
Again the 
transverse branch is weighted by an additional factor $q^{2}$ (compared to
the longitudinal contribution: see equs. (\ref{8}) and (\ref{9})) which 
results 
in an enhancement of the corresponding time--like counterpart of the 
quasielastic peak. 

From figs.\,1a and 2a it is easily read off that approximations which neglect
vacuum polarization suffer from the deficiency that the response function is
no more positive definite over the whole energy range. The sum of vacuum
and density dependent contributions (figs.\,3a and 4a)  however is always 
positive
\footnote{If one compares figs.\,1a and 2a with figs.\,3a and 4a it seems 
that the 
(negative) time--like density dependent contributions start at 
$q_{0} \sim 4\,k_{\mathrm F}$ which is about $1\,k_{\mathrm F}$ below
the threshold for pair production. But one has to keep in mind, that these
curves show the result of two different approximations. As was pointed out 
in section IV the effective nucleon mass is higher when vacuum polarization
is taken into account properly. Consequently the 
${\mathrm N\bar N}$-threshold
${\mathrm min}(2\epsilon_{|\vec q|/2},
\epsilon_{\vec k_{\mathrm F} - |\vec q|}+\epsilon_{k_{\mathrm F}})$ is
shifted according to the corresponding mass shift from $m^* = 0.56\,m$ to
$m^* = 0.78\,m$. So in figs.\,3a and 4a the density dependent contributions
start at the same threshold as vacuum polarization}.
As noticed already in sec.\,III, with regard to vacuum polarization 
$\Pi_{vv}^{\mathrm L}$  does not fall off to zero for 
$q_{0} \rightarrow \infty$ but approaches a constant value 
(see equ. (\ref{17})).
As a consequence of our remarks above this holds true also for the
longitudinal response function (see fig.\,3a). Therefore the real part of 
$\Pi_{\mathrm L}$ in fig.\,3b
cannot be obtained directly from the curve in fig.\,3a making use of 
(\ref{16}) as it was the case in fig \,1. Instead, according to the first
equation of (\ref{18}) one subtraction has to be performed. The meaning of
this renormalization procedure becomes more obvious, if one splits up the
vacuum contribution in fig.\,3a in a constant part (which starts at the 
${\mathrm N\bar N}$--threshold) and a contribution which has an asymptotic behaviour like 
$1/q_{0}^{2}$ for $q_{0} \rightarrow \infty$ (compare equ. (\ref{12})). 
More specifically we write 
${\mathrm R}_{\mathrm L}(q_{0}^{2},\vec q^{\,2})$ in the form:
\begin{equation}\label{23a}
{\mathrm R}_{\mathrm L}(q_{0}^{2},\vec q^{\,2}) = {\mathrm R}_{\mathrm L}(q_{0}^{2}=\infty 
,\vec q^{\,2})\,\Theta(q^{2}-4m^{*\,2}) + 
\Delta {\mathrm R}_{\mathrm L}(q_{0}^{2},\vec q^{\,2})
\end{equation}   
where by construction $\Delta {\mathrm R}_{\mathrm L}(q_{0}^{2},\vec q^{\,2})$ provides
a finite contribution to the integral (\ref{16}) which is harmless. 
The leading order high energy contribution to ${\mathrm Re}\,\Pi_{\mathrm L}$ comes 
from the
$\Theta$--function which may be integrated performing one subtraction:
\begin{equation}\label{23b}
{\mathrm Re}\,\Pi_{\mathrm L}(q_{0}^{2}, \vec q^{\,2})
\sim - q_{0}^{2}\;{\large P}\int \limits_{4m^{*\,2}+\vec q^{\,2}}^{\infty}
\frac{dq_{0}^{'\,2}}{\pi} \frac{1}{q_{0}^{'\,2}(q_{0}^{'\,2}-q_{0}^{2})}
\sim \ln(q_{0}^{2})\;.
\end{equation}   
The negative sign of the principal value integral in (\ref{23b}) follows
from the definition (\ref{22}) of the response function: ${\mathrm R}_{\mathrm L} \sim
- {\mathrm Im}\,\Pi_{\mathrm L} > 0$. In this way we obtain the
same result as by using dimensional regularization \cite{Kurasawa2}, 
\cite{Wehrberger2}.

Because of (\ref{23}) two subtractions are needed in order to calculate
${\mathrm Re}\,\Pi_{\mathrm T}(q_{0}^{2}, \vec q^{\,2})$ from the response function
in fig.\,4a (i.e. one has to integrate 
${\mathrm R}_{\mathrm T}(q_{0}^{2},\vec q^{\,2})/q_{0}^{4}$ and multiply the result
with $q_{0}^{4}$). Analogous to (\ref{23b}) one obtains to leading order in
$q_{0}^{2}$:
\begin{equation}\label{23c}
{\mathrm Re}\,\Pi_{\mathrm T}(q_{0}^{2}, \vec q^{\,2})
\sim q_{0}^{2}\ln(q_{0}^{2})\;.
\end{equation} 
This result can be easily checked in fig.\,4b. Hence the relation (\ref{23})
is valid also for the real parts of $\Pi_{\mathrm L,T}$, i.e. in the high 
energy limit we get the final relation:
\begin{equation}\label{23d}
\Pi_{\mathrm T}(q_{0}^{2}, \vec q^{\,2})
\sim q^{2}\Pi_{\mathrm L}(q_{0}^{2}, \vec q^{\,2})  
\end{equation} 
which will be used in the following
\footnote{Here of course we recover the tensor structure: 
$\Pi^{\mu \nu}_{vac} = [g^{\mu \nu}- q^{\mu}q^{\nu}/q^2]\Pi_{vac}(q^2)$.
So equ. (\ref{23d}) proves that our renormalization procedure preserves
current conservation.}.

RPA-correlations lead to a dramatical decrease of ${\mathrm N\bar N}$ pair production
by a factor $1/2$. In our approximation of point--like nucleons without
anomalous magnetic moment this suppression is a consequence of isospin 
symmetry. For the transverse response function this can be read off 
most easily from equ. (\ref{4}). Because the photon in our approximation couples 
only to protons while meson exchange is charge-independent, the 
`electromagnetic bubbles' $\Pi_{vv}^{\mathrm T}$ for single
particle-hole pair creation as well as in the numerator of the 
RPA-correction are weighted by a factor $1/2$ in contrast to the `mesonic'
bubble in the denominator. For high energies the $q^{2}$-dependent part
of this denominator becomes much larger than 1. This can be checked 
substituting 
$\Pi_{vv}^{\mathrm T}$ according to (\ref{23d}): the leading factor $q^{2}$ 
cancels the $1/q^{2}$-decrease of the propagator $G_{v}(q^{2})$, so that 
the RPA-denominator in (\ref{4}) goes like 
$(1-g_{v}^{2}\,\Pi_{vv}^{\mathrm L}) \sim 1 - g_{v}^{2}\ln(q_{0}^{2})$ (see 
equ. (\ref{23b})). Hence for high 
energies the 1 in the RPA-denominator may be neglected.
Then the rest of the denominator in (\ref{4}) cancels against the 
corresponding part in the
nominator so that (because of the isospin-factor 1/2) one gets from 
the RPA-correction a contribution which is exactly (-1/2) times the
single ${\mathrm N\bar N}$-bubble contribution
\footnote{Note that the starting energy of this asymptotic behaviour depends
on the numerical value of $g_{v}^{2}$ which is a parameter of the model.
If $g_{v}^{2} >> 1$ (compare equ. (\ref{21})) the reduction by a factor 1/2
starts practically already at threshold which is confirmed by figs.\,3 and 4.}.
The same holds true for the longitudinal part (\ref{3}) of the polarization
propagator. As pointed out in the section preceding equ. (\ref{13}), 
$\Pi_{sv}$ is a purely density dependent contribution. From equ. (\ref{10})
it follows (and can be easily checked in figs.\,1 and 2) that 
$\Pi_{sv}(q_{0},\vec q^{\,2}) = 0$ for 
$q_{0} > \epsilon_{\vec k_{\mathrm F} + |\vec q|}+\epsilon_{k_{\mathrm F}}$.
Consequently only the third term in the nominator and the first term
in the denominator of (\ref{3}) are nonzero for $q_{0}$ high enough. So in this 
limes $\Pi_{{\mathrm RPA}}^{\mathrm L}$ can be written:
\begin{equation}\label{23e}
\left[\Pi_{{\mathrm RPA}}\right]_{\mathrm L}(q_{0},\vec q^{\,2}))\;
=\; \frac{e^{2}}{2}\Pi_{vv}^{\mathrm L}(q_{0},\vec q^{\,2}) 
\,+\, e^{2}\, \frac{\frac{q^{2}}{\vec q^{\,2}}G_{v}
(\frac{1}{2}\Pi_{vv}^{\mathrm L})^{2}}
{(1-\frac{q^{2}}{\vec q^{\,2}}G_{v}\Pi_{vv}^{\mathrm L})} \;.
\end{equation}
Completely analogous to the RPA-correction to 
$\Pi_{\mathrm T}$, the 1 in the denominator of (\ref{23e}) may be neglected
and one gets a negative contribution of just one half of the 
single pair production amplitude.

This argumentation on the one hand is very general because it is based
only on the isospin symmetry of the NN-interaction and holds true
beyond the $\sigma\omega$--model. So for point--like
nucleons we would expect a suppression of the cross section $e^{+}e^{-}
\rightarrow {\mathrm p\bar p}$ by 50\%, inside nuclear matter as well as in the
vacuum. On the other hand we know that nucleons are {\em not point--like} 
particles, which may invalidate (at least partly) our reasoning. 
Nevertheless, we think that our model calculation has shown that many
body correlations can produce effects far beyond the usually assumed level
of 10-20\%. This result should not alter dramatically taking the finite 
size of nucleons properly into account.

An observable which would be directly affected by such many body effects
is the ratio:
\begin{equation}\label{24}
{\mathrm R} = \frac{\sigma(e^{+}e^{-} \rightarrow {\mathrm hadrons})}
{\sigma(e^{+}e^{-} \rightarrow \mu^{+}\mu^{-})}
\end{equation}
Within the simple quark--parton model (QPM) -- which corresponds to the one 
bubble approximation in our calculations -- R is given by:
\begin{equation}\label{25}
{\mathrm R} = 3\, \sum_{q} e_{q}^{2}\; ,
\end{equation}
where the index $q$ runs over all quark flavours with charge $e_{q}$.
A review on the value of R measured in several experiments can be found
in \cite{Marshall}. Unfortunately our considerations concerning the process
$e^{+}e^{-}\rightarrow {\mathrm p \bar p}$ cover only a fraction of R which 
makes it difficult to compare directly with these experiments. 
Nevertheless, a thorough 
analysis of the above general isospin argument on the quark level seems
to be worthwhile but goes beyond the scope of the present paper.

\section{COLLECTIVE PHENOMENA}
Another remarkable difference between the full and dashed curves in figs.\,1-4 
are the narrow peaks in the time--like
energy regime below the threshold for pair production. These belong to
the collective excitations already discussed in sections I and II. 
The correct structure of these poles in the real {\em and} imaginary part
of the polarization propagator may be obtained only if the calculation is
extended to complex energies. The sharp $\delta$--peaks in the 
response functions (see part a of figs. 1-4) are the result of an analytic
continuation back to the real energy axis according to (\ref{22}).
For ${\mathrm Im}(z) = 0$ the 'one bubble' contribution to the response
function is zero in the unphysical region $|\vec q| \le q_{0} \le
{\mathrm min}(2\epsilon_{|\vec q|/2},
\epsilon_{\vec k_{\mathrm F} - |\vec q|}+\epsilon_{k_{\mathrm F}})$
\footnote{There are no single particle excitations in this energy range}. 
As a consequence, also the nominator of the RPA--corrections in equs. (\ref{3})
and (\ref{4}) vanishes for these energies. Hence zeros of the denominator,
which are the signal of collective excitations, lead to an indefinite 
expression for the RPA--formulas. 

Former calculations of the electronuclear response functions 
\cite{Kurasawa1}--\cite{Horowitz2} were restricted only to real spacelike
energies (more specific: to the quasi elastic bump) and therefore did not 
have to care about this problem.
On the other hand, the authors who looked for timelike excitations \cite{Chin},
\cite{Hatsuda}, \cite{Lim}, \cite{Jean} only calculated the denominators
of equs. (\ref{3}) and (\ref{4}) and looked for zeroes in this quantity.
As will be shown below, for real excitations this method provides
the correct spectrum, but in the case of an instability such a treatment
may be misleading.

From counting of degrees of freedom (see section I) one would expect
two peaks in the longitudinal and one (twofold degenerate) peak in the 
transverse response function. But figs.\,1 and 2, where only density dependent 
contributions are taken into account, show one single excitation in the 
longitudinal and none at all in the transverse branch. However, including
vacuum polarization the curves in figs.\,3 and 4 show the correct number of
peaks. 

Here it is instructive to look at the RPA-denominators in both approximations
which are shown in fig.\,5 for the longitudinal branch (the following arguments
hold true also for the transverse response). The large peaks in the real parts
of both figures come from the free $\sigma$ and $\omega$ propagator in the
denominator of equ. (\ref{3}) with masses given in equ. (\ref{21}). The dispersion
of the dressed mesons inside nuclear matter is determined by the zeroes of the
full curves in fig.\,5 (but, as said before, only if the imaginary part, i.e. the
dashed curve in fig.\,5, vanishes too). Including vacuum polarization (fig.\,5b)
we find a reduction of the effective meson masses because the zeroes are situated
below the peaks of the free propagators. This result is in accordance with 
previous calculations \cite{Shiomi}, \cite{Hatsuda}, \cite{Jean}. On the contrary, neglecting
Dirac sea effects, the real zeroes of the RPA--denominator (from which only the 
first corresponds to a well defined excitation)  are situated above
the free meson poles which means an enhanced mass of the quasiparticles.

In order to understand the reason for this difference in the two approximations
it is helpful to take a closer look at the shape of the peaks in fig.\,5. If one compares 
fig.\,5a with fig.\,5b it is easy to verify a different sign of the meson poles. While
in fig.\,5a the real part of the denominator is positive below the poles and negative
above, the zeroes corresponding to the quasiparticles also lie above the poles. 
Exactly opposite is the situation for the calculation including vacuum polarization
(fig.\,5b) which leads to the reduction of the meson mass. 

Now, the mechanism {\em how} Dirac sea effects cause this negative mass shift
becomes clearer, if one looks at the denominator of the RPA correction in
equ. (\ref{3}). In the energy range between the photon point $q^{2} = 0$ and
the threshold for pair production $q^{2} = m^{*\,2}$ the mixing term of scalar
and vector degrees of freedom $\Pi_{sv}^{2}$ can be neglected in comparison
with the pure terms $\Pi_{vv}^{\mathrm L}$ and $\Pi_{ss}$. So for these energies
the RPA denominator may be approximated by the expression
$(1-\frac{q^{2}}{\vec q^{\,2}}G_{v}\Pi_{vv}^{\mathrm L})(1-G_{s}\Pi_{ss})$.
The poles of the propagators $G_{s}$ and $G_{v}$ provide the peaks discussed
above, the signs of which depend on the sign of $\Pi_{vv}^{\mathrm L}$
and $\Pi_{ss}$ in the neighbourhood of these poles respectively. Because without 
RPA corrections $\Pi_{vv}^{\mathrm L}$ is directly proportional to the 
longitudinal response function (see the first term in equ. (\ref{3})) its behaviour 
can be red off from the dashed curves in figs.\,1 and 3. For the energies
under consideration the imaginary parts vanish. Neglecting the Dirac sea,
the dashed curve in fig.\,1b for $\vec q^{\,2} < q_{0}^{2} < \vec q^{\,2} + m^{*\,2}$
has positive sign. Because the scalar and vector meson masses are well separated,
near the vector pole $(1-G_{s}\Pi_{ss}) \sim 1 > 0$. This leads to the peak
structure described above and an enhancement of the $\omega$ mass inside nuclear
matter in this approximation.

Comparing fig.\,1b with fig.\,3b it is easy to see, that vacuum polarization
provides a {\em negative} contribution to the real part of $\Pi_{vv}^{\mathrm L}$.
The result of this is the switch of sign in the free $\omega$ pole observed in
fig.\,5b with the consequence of a reduced mass of the dressed quasiparticle. 
According to equ. (\ref{18}) the magnitude of the negative Dirac sea contribution and
thus also the meson mass shift depends on the amplitude for nucleon--antinucleon pair 
production. The threshold for this process inside nuclear matter is determined 
by the effective nucleon mass $m^{*}$ which is the free nucleon mass $m$ reduced
by a term proportional to the scalar density $<\bar \psi \psi>$ \cite{Serot}. 
Hence analogous to the results of the QCD inspired approach 
\cite{Li}, \cite{Hatsuda} with a meson mass shift due to medium modifications
of the quark condensate $<\bar u u>$ also in the Walecaka model a scalar
density is responsible for the changes in the dispersion of the mesonic degrees
of freedom.

Despite the preceding considerations, 
%on the basis of fig.\5 
as an illustration of our method, gave us a qualitative 
picture of the underlying mechanism for the meson mass shift, it is not the purpose 
of the present paper to analyze this effect in more detail. Rather it is our goal to
point out the importance of vacuum polarization for the stability of the nuclear
matter system. In \cite{Jean} Jean et. al. argue that,
neglecting vacuum polarization, the zero in the real part
of fig.\,5a at $\omega/k_{\mathrm F} {\sim} 4.5$ belongs to a collective excitation
with "an unphysical (i.e. negative) decay width" because of the nonzero negative
imaginary part of the RPA-denominator in this energy range. However, according 
to figs.\,1 and 2 for {\em real} energies the response functions itself do not show 
any sign of such a unphysical mode. The cusp like structures at 
$\omega/k_{\mathrm F} {\sim} 4$ 
in the real parts of the longitudinal and transverse matrix elements of the 
polarization tensor (figs.\,1b and 2b) are already present in the one bubble
approximation and therefore are not the result of collective phenomena.

The true situation unfolds itself only if the calculation of the response functions
is extended to complex energies (see figs.\,6 and 7). The peaks which are missed on 
the real energy axes reappear as {\em undamped} poles in the upper complex half 
plane. So in contrast to the prediction of \cite{Jean} an approximation which
neglects vacuum polarization does not lead to a negative decay width of the 
$\omega$ meson but to collective excitations which are unphysical because, according 
to the discussion in section II, they manifestly destroy causality. Consequently 
calculation within this approximation cannot provide reliable results.

\section{CONCLUSIONS}
We studied qualitative properties of the electro--nuclear response functions
for nuclear matter with respect to single particle and collective excitations.
Performing the calculations exemplary in Walecka's $\sigma\omega$--model, 
the main results were obtained from general arguments which should hold true
far beyond this simple model for the medium NN--interaction.

In contrast to former calculations where the main attention was concentrated
on the quasielastic region, we extended our considerations \,\,1) to the 
time--like energy regime and \,\,2) to complex energies. The latter was
achieved making use of dispersion relations. As an alternative to 
dimensional regularization we introduced a renormalization scheme which
is more appropriate to our approach: infinite vacuum contributions were
removed performing suitable subtractions in the dispersion integrals.
With the appropriate boundary conditions this method leads to the same results
as dimensional regularization.

Neglecting vacuum polarization we found that there is a violation of causality
because of unstable longitudinal and transverse collective excitations in 
the upper complex energy plane. Including vacuum polarization, these 
collective modes are shifted to the real axes and change to proper excitations 
of the system. So we have shown that vacuum polarization is crucial for the 
stability of nuclear matter.

In addition we have shown that RPA-corrections reduce the production
amplitude of $p\bar p$ pairs by a factor 1/2. This result is remarkable
because it holds true for any interaction between point--like nucleons
which preserves isospin symmetry. Our argument may have interesting 
consequences for the process $e^{+}e^{-}\rightarrow {\mathrm p\bar p}$.
\begin{figure}
\vspace{9in}
\centerline{
\psfig{figure=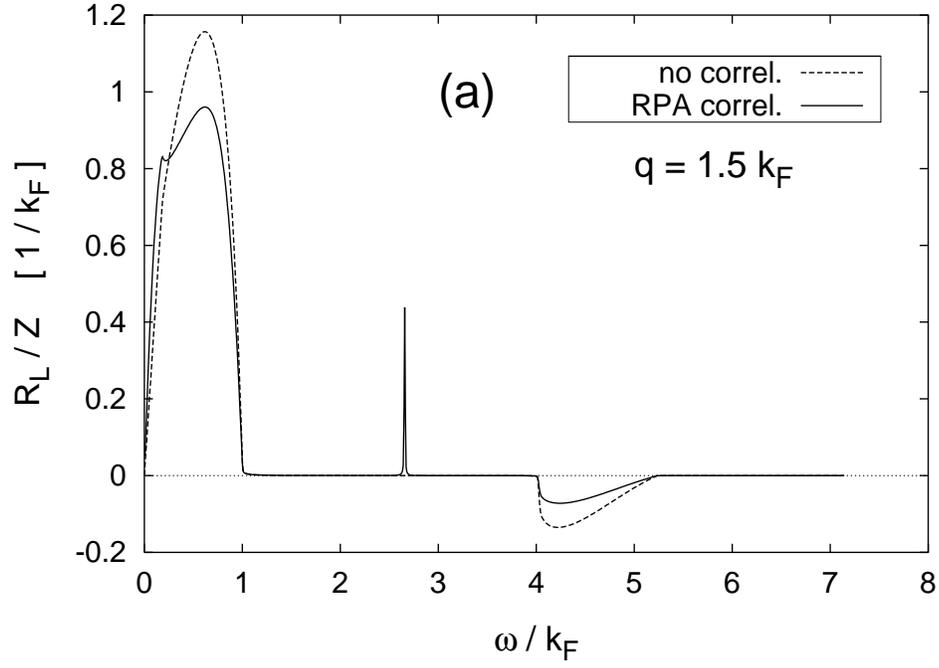,height=3.6in,width=5.1in,angle=-90}}
\vspace{0.5in}
\centerline{
\psfig{figure=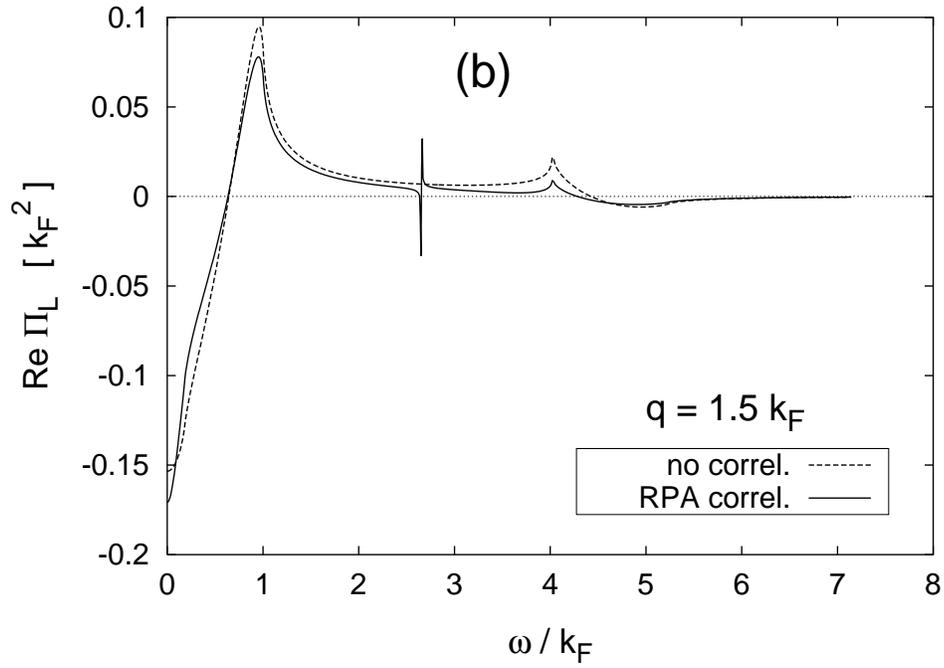,height=3.6in,width=5.1in,angle=-90}}
\vspace{.2in}
\caption[]{\label{fig1}{\small
(a) Longitudinal response function and (b) real part of the longitudinal 
matrix 
element $\Pi_{\mathrm L}$ of the polarization tensor for nuclear matter 
at a momentum transfer of $1.5\,k_{\mathrm{F}}$ without vacuum 
polarization. The dashed curve shows a calculation for space--like and 
time--like energies neglecting 
RPA--correlations, while these are included in the full curve.}}
\end{figure}
\newpage
\begin{figure}
\vspace{9in}
\centerline{
\psfig{figure=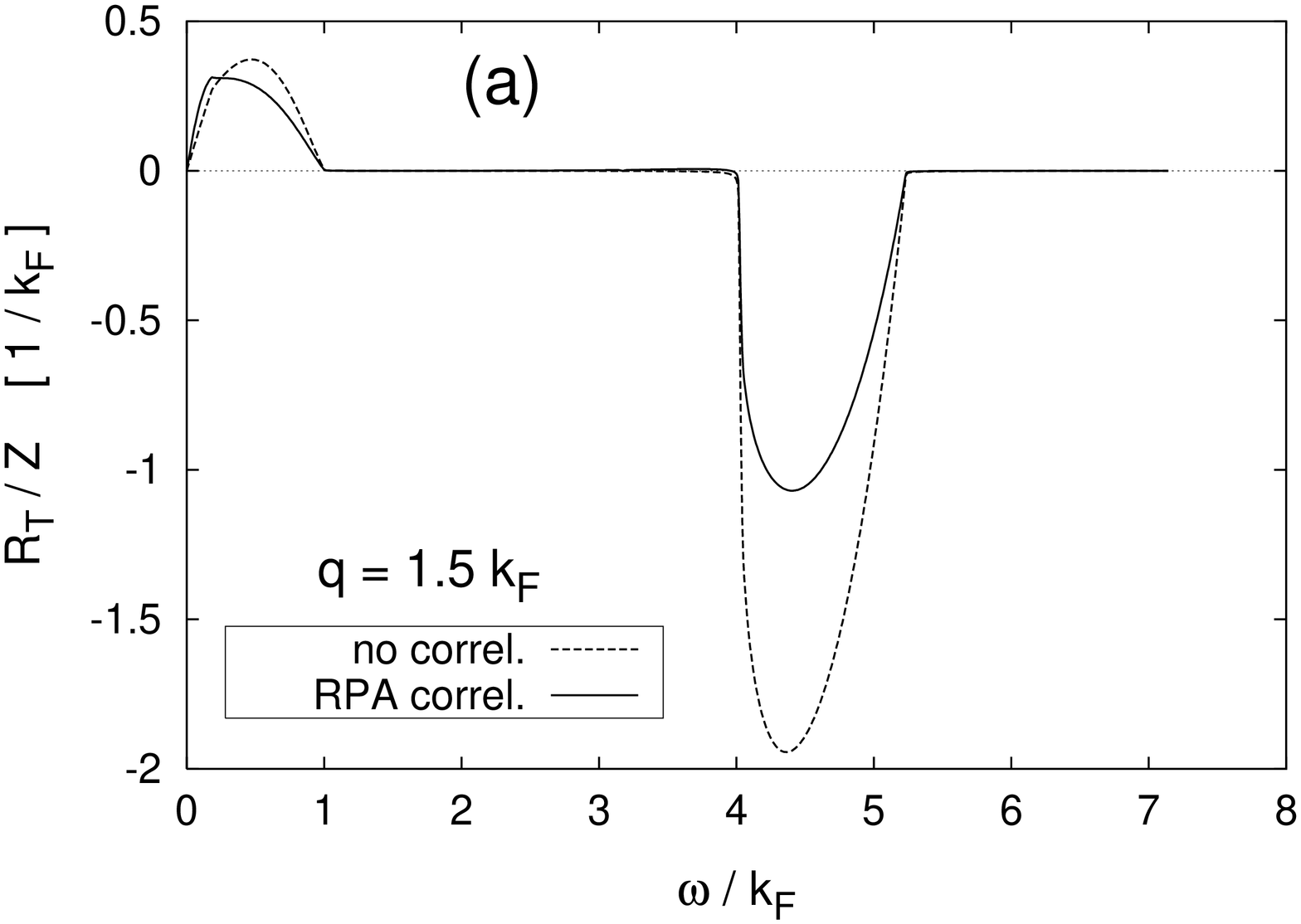,height=3.6in,width=5.1in,angle=-90}}
\vspace{0.5in}
\centerline{
\psfig{figure=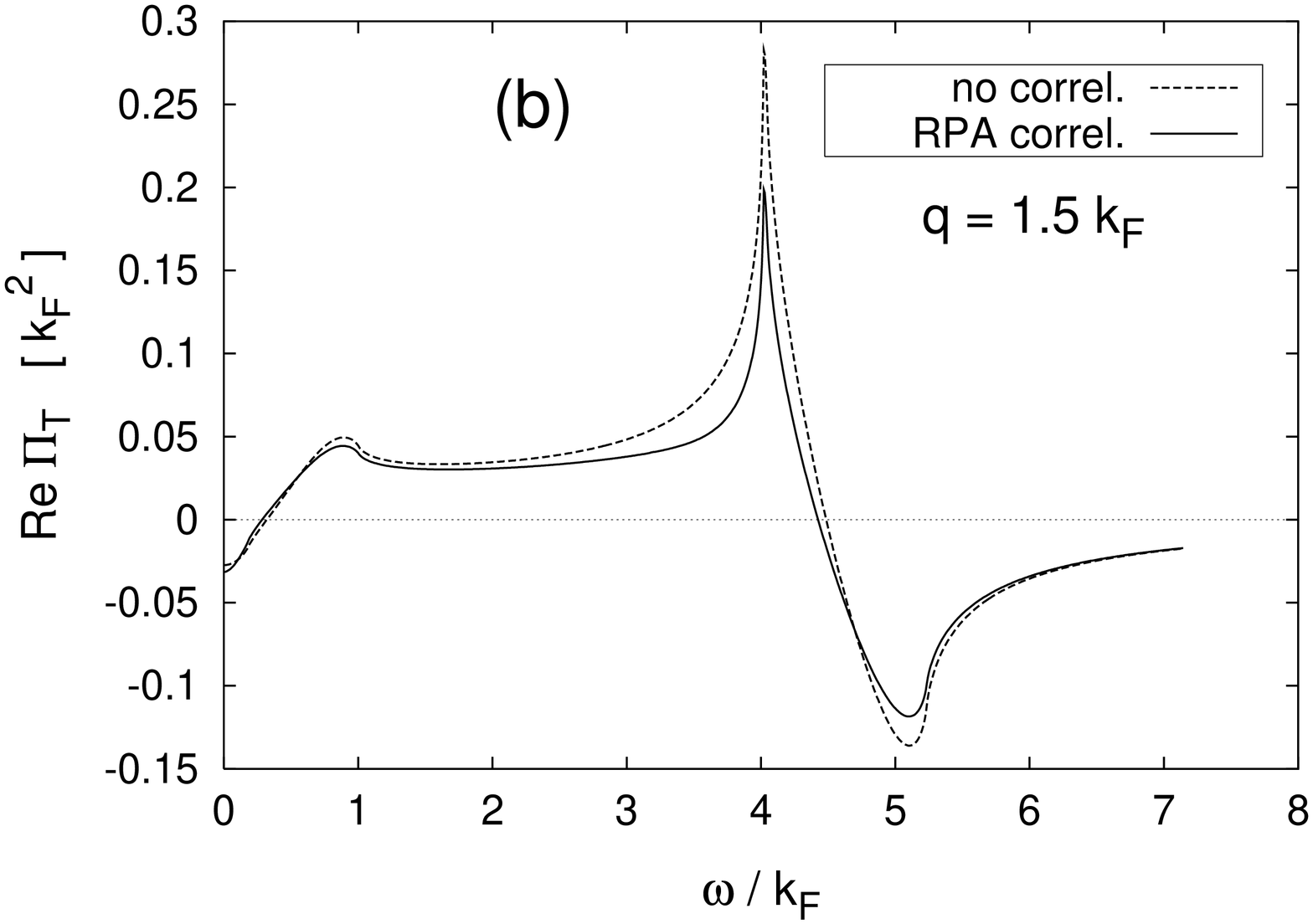,height=3.6in,width=5.1in,angle=-90}}
\vspace{.2in}
\caption[]{\label{fig2}{\small
Same as fig.\,1 for
(a) the transverse response function and (b) the real part of the transverse 
matrix 
element $\Pi_{\mathrm T}$ of the polarization tensor.}}
\end{figure}
\newpage
\begin{figure}
\vspace{9in}
\centerline{
\psfig{figure=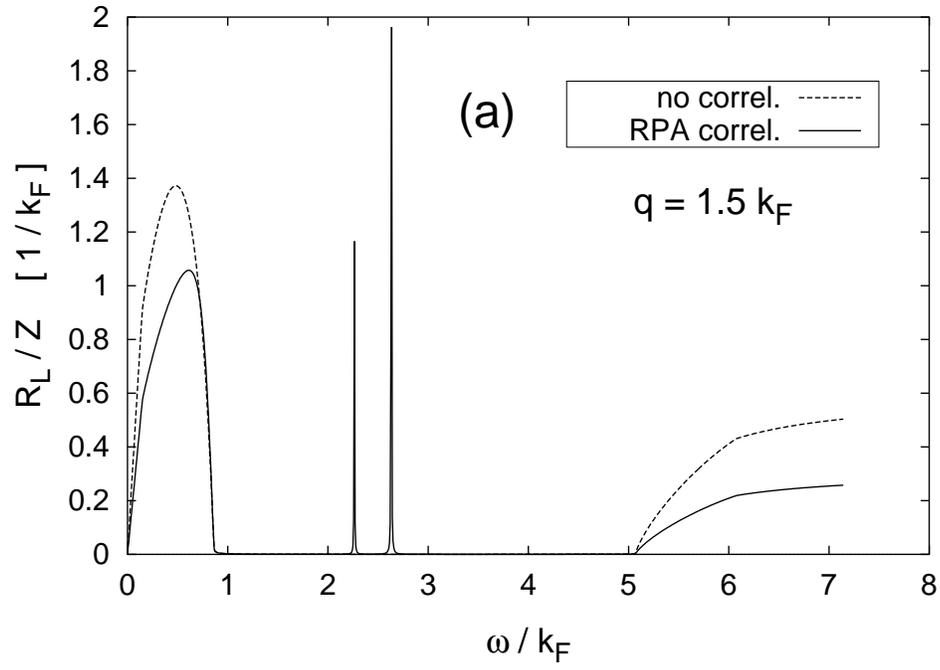,height=3.6in,width=5.1in,angle=-90}}
\vspace{0.5in}
\centerline{
\psfig{figure=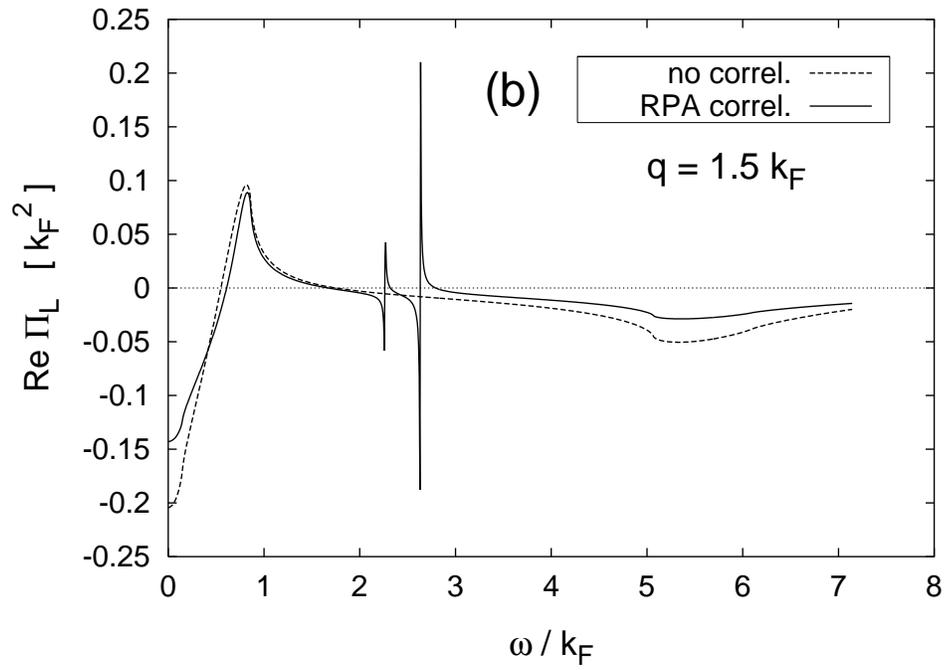,height=3.6in,width=5.1in,angle=-90}}
\vspace{.2in}
\caption[]{\label{fig3}{\small
Same as fig.\,1 including vacuum polarization.}}
\end{figure}
\newpage
\begin{figure}
\vspace{9in}
\centerline{
\psfig{figure=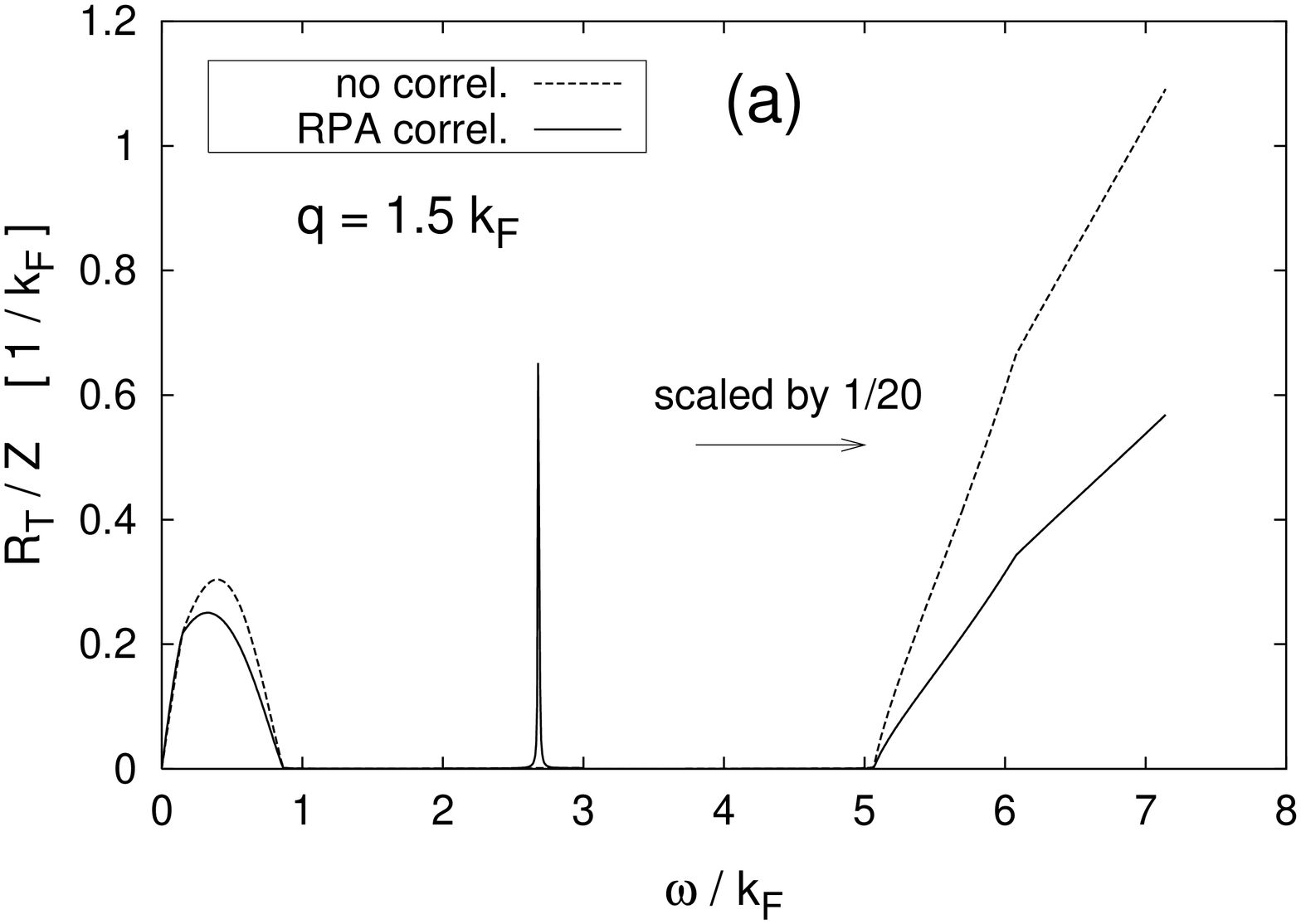,height=3.6in,width=5.1in,angle=-90}}
\vspace{0.5in}
\centerline{
\psfig{figure=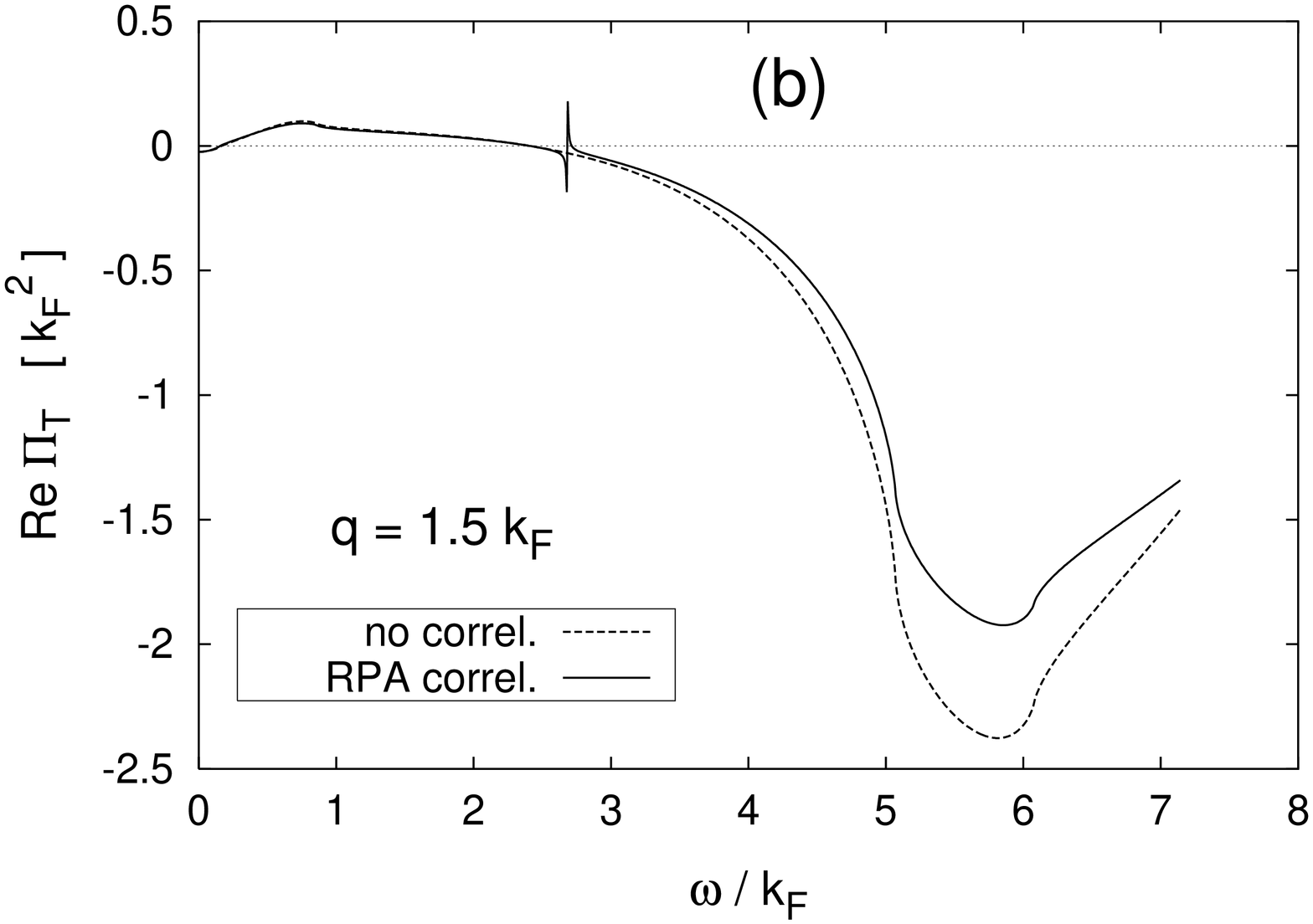,height=3.6in,width=5.1in,angle=-90}}
\vspace{.2in}
\caption[]{\label{fig4}{\small
Same as fig.\,2 including vacuum polarization. Note that in (a) because
of the fast increase of the transverse vacuum polarization this part of the
curves is scaled by 1/20.}}
\end{figure}
\newpage
\begin{figure}
\vspace{9in}
\centerline{
\psfig{figure=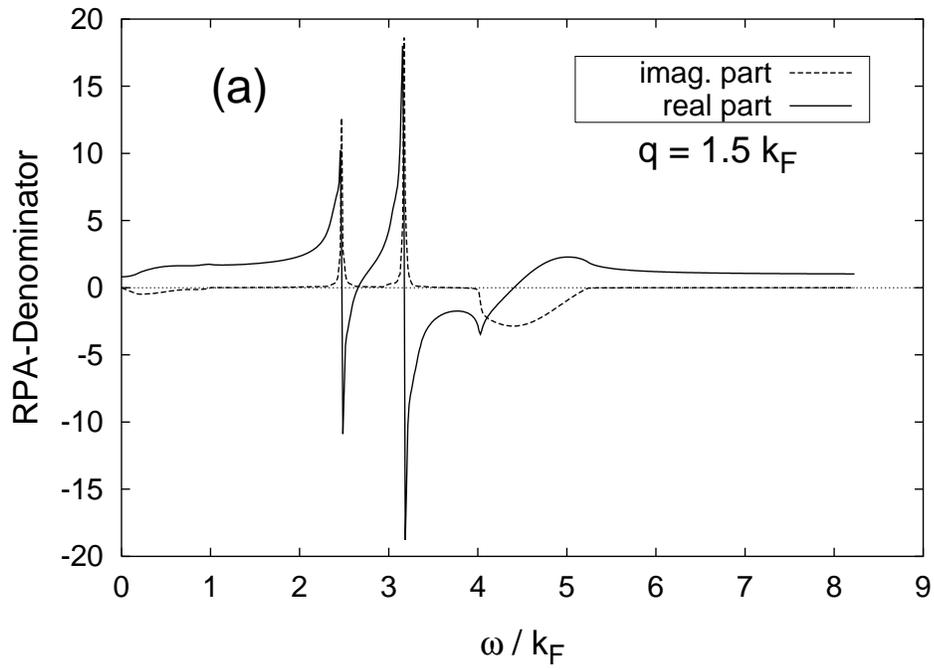,height=3.6in,width=5.1in,angle=-90}}
\vspace{0.5in}
\centerline{
\psfig{figure=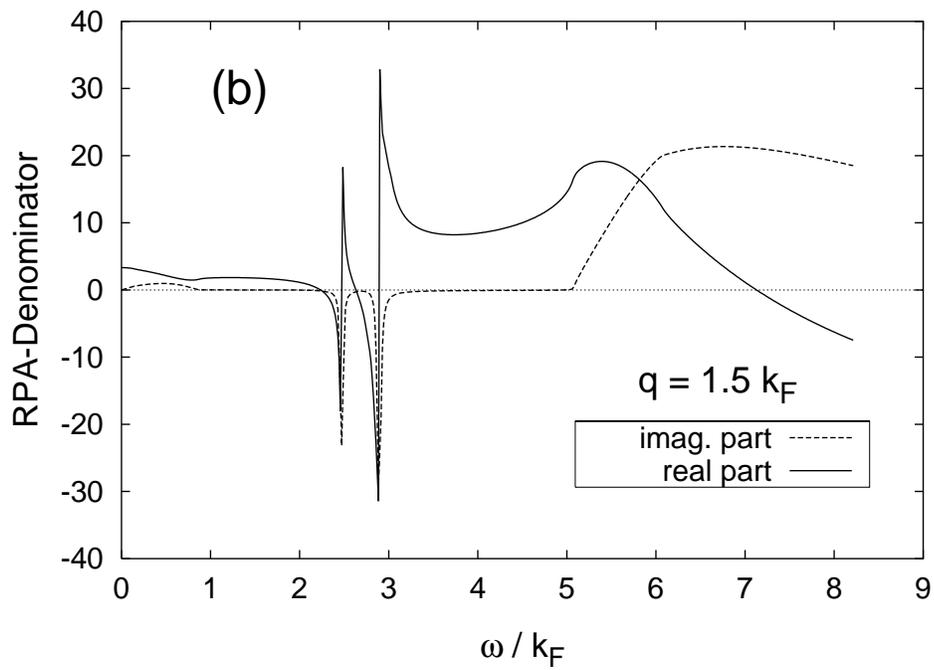,height=3.6in,width=5.1in,angle=-90}}
\vspace{.2in}
\caption[]{\label{fig5}{\small
Real part (full curves) and imaginary part (dashed) of the longitudinal
RPA-denominator according to equ. (\ref{3}). While in (a) vacuum 
polarization is neglected in (b) these contributions are included properly.}}
\end{figure}
\newpage
\begin{figure}
\vspace{9in}
\centerline{
\psfig{figure=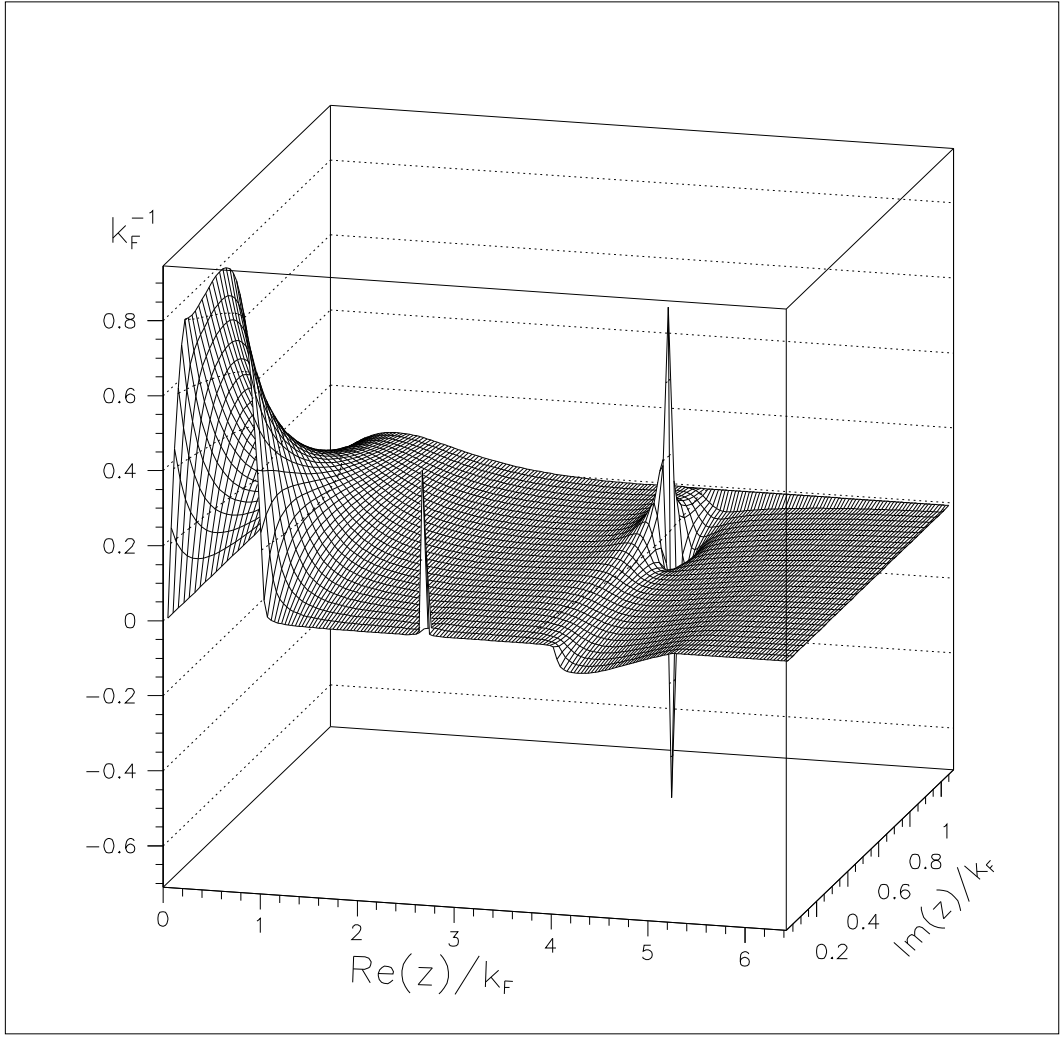,height=7in,width=5in}}
\vspace{.2in}
\caption[]{\label{fig6}{\small
Extension of the longitudinal RPA--response function to complex energies 
neglecting vacuum polarization (compare the full curve in fig.\,1a). The big 
singularity in the 
upper complex energy plane is the signal for the violation of causality in  
this approximation.}}
\end{figure}
\newpage
\begin{figure}
\vspace{9in}
\centerline{
\psfig{figure=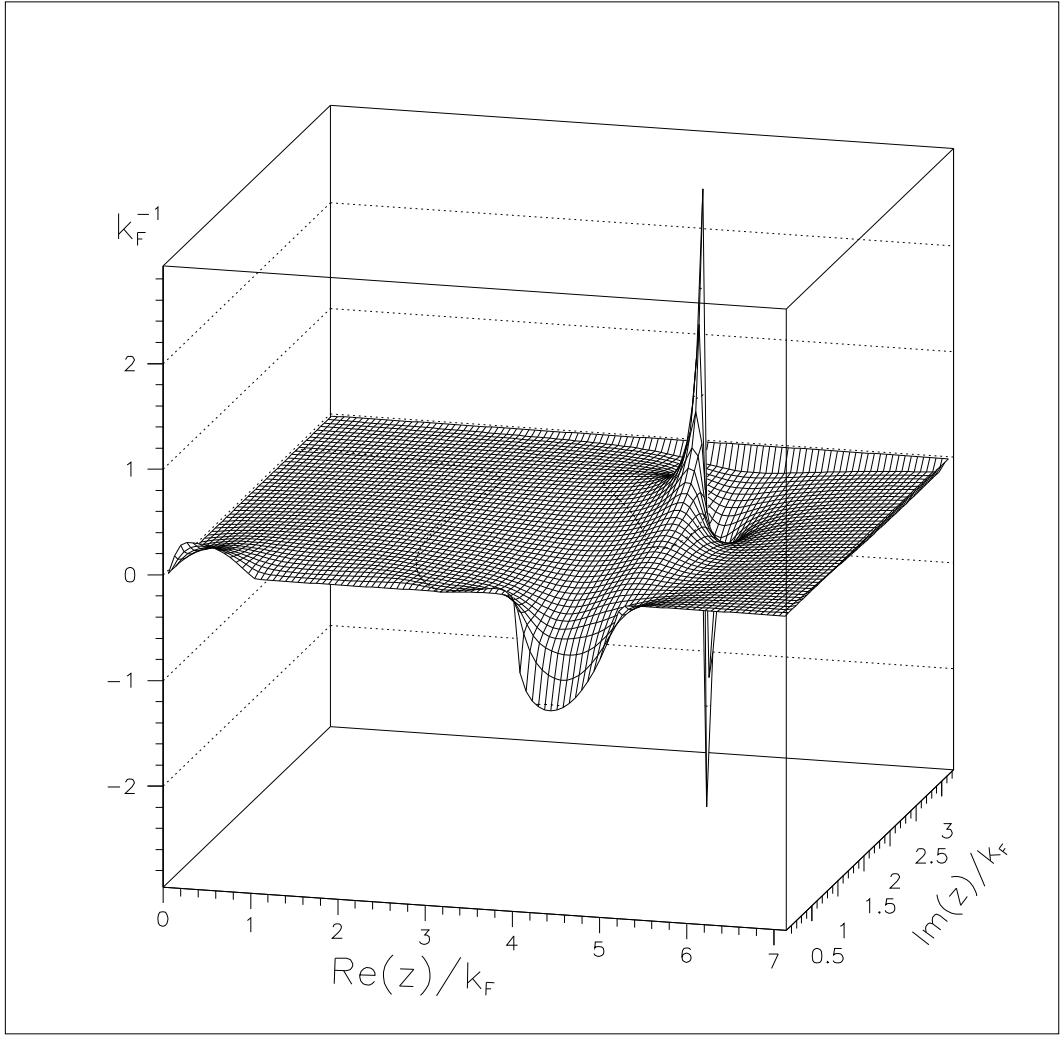,height=7in,width=5in}}
\vspace{.2in}
\caption[]{\label{fig7}{\small
Same as fig.\,6 for the transverse response function.}}
\end{figure}
\end{document}